\documentclass[useAMS,usegraphicx,usenatbib]{mn2e}   \usepackage{times}
\usepackage{rotate} \usepackage{amssymb}
\title[Morphologies   of  X-ray   selected   AGN  hosts]{Host   galaxy
  morphologies of  X-ray selected  AGN: assessing the  significance of
  different black hole fueling  mechanisms to the accretion density of
  the  Universe  at  z$\sim$1.}   \author[Georgakakis et  al.   ]  {A.
  Georgakakis$^{1}$\thanks{email: age@astro.noa.gr}, A.  L.  Coil$^2$,
  E.   S.  Laird$^3$,  R.  L.   Griffith$^4$, K.   Nandra$^3$,  J.  M.
  Lotz$^{5}$,\\\\ {\rm \LARGE C.   M. Pierce$^{6}$, M.  C. Cooper$^7$,
    J.  A.   Newman$^8$, A.   M.   Koekemoer$^9$}  \\ \\  $^1$National
  Observatory   of   Athens,   V.   Paulou  \&   I.   Metaxa,   11532,
  Greece\\ $^2$Department  of Physics and Center  for Astrophysics and
  Space  Sciences, University  of California,  San Diego,  9500 Gilman
  Dr.,   La  Jolla,  CA   92093\\  $^3$Astrophysics   Group,  Blackett
  Laboratory, Imperial  College, Prince Consort  Rd , London  SW7 2AZ,
  UK\\   $^4$Jet  Propulsion   Laboratory,  California   Institure  of
  Technology,  Pasadena,  CA  91109\\ $^5$National  Optical  Astronomy
  Observatory,   950   North   Cherry   Avenue,  Tucson,   AZ   85719,
  USA\\  $^6$Department of  Physics, University  of  California, Santa
  Cruz,  1156 High  Street, CA  95064, USA\\  $^7$Steward Observatory,
  University of  Arizona, 933 N.  Cherry Ave., Tucson,  AZ 85721-0065,
  USA\\ $^8$University of Pittsburgh, Physics \& Astronomy Department,
  3941 O'Hara Street Pittsburgh,  PA, 15260, USA\\ $^9$Space Telescope
  Science Institute, 3700 San Martin Drive, Baltimore, MD 21218, USA }
\begin{document}
\maketitle

\begin{abstract}  We use morphological  information of  X-ray selected
AGN hosts  to set limits on  the fraction of the  accretion density of
the Universe at  $z\approx1$ that is not likely  to be associated with
major  mergers.   Deep  X-ray  observations  are  combined  with  high
resolution optical data from the  Hubble Space Telescope in the AEGIS,
GOODS  North  and GOODS  South  fields  to  explore the  morphological
breakdown of  X-ray sources in the redshift  interval $0.5<z<1.3$. The
sample  is  split  into  disks, early-type  bulge-dominated  galaxies,
peculiar  systems  and  point-sources  in  which  the  nuclear  source
outshines  the  host  galaxy.    The  X-ray  luminosity  function  and
luminosity  density of  AGN at  $z\approx1$  are then  calculated as  a
function  of morphological  type.  We  find that  disk-dominated hosts
contribute  $30\pm9$ per  cent  to  the total  AGN  space density  and
$23\pm6$ per cent to the  luminosity density at $z\approx1$.  We argue
that AGN  in disk galaxies are  most likely fueled {\it  not} by major
merger events but by minor interactions or internal instabilities.  We
find evidence that these mechanisms may be more efficient in producing
luminous  AGN ($L_X>10^{44}  \, \rm  erg \,  s^{-1}$) compared  to
predictions for the stochastic fueling of
massive black holes in disk galaxies.
\end{abstract}
\begin{keywords}   Surveys  --   galaxies:  starbursts   --  galaxies:
evolution -- X-rays: galaxies
\end{keywords}

\section{Introduction}\label{sec_intro}

It is  well established that  Active Galactic Nuclei (AGN),  which are
signposts  of  accretion  events  onto the  supermassive  black  holes
(SMBHs) at the centres of  galaxies, evolve strongly with cosmic time.
Observational  programs over  the last  35 years  have shown  that the
space  density of  these systems  has  a broad  peak at  $z\approx1-3$
\citep[e.g.][]{Hasinger2005,Brusa2008,Silverman2008_XLF,Aird2008}
followed  by   a  decline  by   almost  2\,dex  to  the   present  day
\citep[e.g.][]{Ueda2003,  Hasinger2005,   Barger2005}.   The  physical
interpretation of  the observed rapid evolution of  the AGN population
however, is still a matter of considerable debate.  An important piece
of the puzzle  is the process that triggers  the accretion of material
onto the central SMBH.

A major galaxy merger is one of the popular mechanisms suggested to be
responsible  for  fueling   AGN.   Numerical  SPH  (Smoothed  Particle
Hydrodynamic)  simulations demonstrate that  these violent  events are
very  efficient  in  funneling  gas  to  the  nuclear  galaxy  regions
\citep[e.g.][]{Hernquist1989,Barnes1991,Barnes1996},  where it  can be
consumed by the SMBH \citep{Springel2005,DiMatteo2005}.  This scenario
is also  attractive because  galaxy bulges are  built parallel  to the
formation  of SMBHs  at their  centres, thereby  providing  a physical
interpretation  for  the  observed  tight  correlation  between  bulge
velocity   dispersion  and   black  hole   mass  in   nearby  galaxies
\citep[e.g.][]{Ferrarese2000,Gebhardt2000}.    As   a   result,   most
semi-analytic    cosmological   simulations   of    galaxy   formation
\citep[e.g.][]{Somerville2008}  use  major   mergers  as  the  primary
mechanism for  growing SMBHs. In this scenario,  changes with redshift
of the  AGN space density are  intimately related to  the evolution of
the major merger rate in the Universe.

Observations provide some support  to the simulations described above,
thereby suggesting that SMBHs can be fuelled by major mergers.  In the
local Universe for example,  the population of Ultra-Luminous Infrared
Galaxies   (ULIRGs),   which    are   dominated   by   major   mergers
\citep[e.g.][]{Surace1998,Borne2000,Farrah2001},   include   a   large
fraction of systems  that show AGN signatures at  optical and/or X-ray
wavelengths   \citep{Sanders1996,Franceschini2003}.   Additionally,  a
large fraction  of luminous QSOs  at low redshift are  associated with
either  morphologically disturbed  galaxies, suggesting  ongoing tidal
encounters  \citep{Canalizo2001,Guyon2006}, or  early type  hosts with
fine structure in their optical light distribution, indicative of past
interactions  \citep[e.g.][]{Canalizo2007,Bennert2008}.  Moreover, the
excess  of optical  neighbours  around $z<0.4$  QSOs  on small  scales
($\approx    0.1-0.5$\,Mpc)   compared    to    $L_{\star}$   galaxies
\citep{Serber2006}  and   the  higher  fraction  of   QSO  pairs  with
separations   $<0.1$\,Mpc  compared  to   the  expectation   on  large
($>3$\,Mpc) scales  \citep{Hennawi2006, Myers2007}, both  suggest that
galaxy mergers play an important role in the evolution of QSOs.

Despite the  evidence above, it is  also recognized that  many AGN are
not fueled by major mergers.  In the nearby Universe for example, many
Seyferts,  which represent  intermediate and  low luminosity  AGN, are
associated with  non-interacting spirals \citep[e.g.][]  {Ho1997}. The
small scale environment ($<$0.1\,Mpc)  of narrow line AGN, including a
large fraction of  Seyferts, in the Sloan Digital  Sky Survey suggests
that major galaxy interactions cannot explain the observed activity in
the bulk  of the population  \citep{Li2008}.  It is believed  that the
SMBHs of  these systems accrete cold gas  stochastically by mechanisms
such as bar instabilities or minor tidal disruptions rather than major
mergers  \citep{Hopkins_Hernquist2006}.   It  is also  suggested  that
minor interactions and secular evolution are important at low redshift
($z\la 1$) and low luminosities (e.g.   $L_X \la 10^{43} \rm \, erg \,
s^{-1}$), while  major mergers dominate  at high redshifts  and bright
luminosities    \citep[e.g.][]{Hasinger2008,   Hopkins_Hernquist2006}.
Compared to models  where major mergers drive the  history of luminous
accretion  at all epochs  \citep[e.g.][]{Hopkins2006, Somerville2008},
the  scenario  above  offers  an  alternative  interpretation  of  the
evolution  of AGN  based on  a change  with redshift  of  the dominant
fueling mode of the SMBH.

Although there is  no doubt that AGN are fueled  by mechanisms that do
not involve  major mergers, as yet  there are no robust  limits on the
relative  contribution  of  such  mechanisms to  the  total  accretion
luminosity.  One way to address  this issue is to study the morphology
of the  AGN host galaxies. In  the standard view of  major mergers any
pre-existing  disks are destroyed  to form  a bulge  dominated remnant
\citep[e.g]{Barnes1996}.     Therefore,   AGN    hosted    by   either
disturbed/irregular   systems  or   spheroidal  galaxies   are  likely
associated  with  ongoing or  past  major  mergers, respectively.   In
contrast, AGN in disk galaxies  are candidates for the stochastic SMBH
fueling  mode.  Although  recent simulations  show that  under certain
conditions (e.g.   high gas fraction) disks can  survive major mergers
\citep[e.g.][]{Springel_Hernquist2005,   Robertson2006,   Hopkins2008,
  Governato2008}, it is  also claimed that the formation  of a remnant
with a large  disk requires conditions which are  often not associated
with  substantial SMBH  fueling \citep[e.g.][]{Hopkins_Hernquist2008}.
The fraction of  AGN in disk galaxy hosts  therefore, remains a strong
constraint on the fueling  mechanism and/or the conditions under which
SMBH growth occurs.

At low  redshift for example, about  30 per cent of  the powerful QSOs
live in  spirals \citep{Canalizo2001, Guyon2006}, while  at $z \approx
1$, close to the peak of  the accretion history of the Universe, about
20-30 per cent of the X-ray  selected AGN are hosted by late-type disk
galaxies \citep{Pierce2007,  Gabor2008}. These observations  are often
interpreted as  examples of SMBHs fueled by  minor interactions and/or
secular  evolution.   What is  uncertain  however,  is the  luminosity
distribution of the  AGN in disk hosts, i.e. what  fraction of the AGN
space and luminosity  density is associated with this  type of galaxy.
This  is essential  to  place a  limit  on the  significance of  minor
interactions and/or internal instabilities to the accretion history of
the Universe.

In this  paper we address this  issue by estimating, as  a function of
host galaxy  morphology, the X-ray  luminosity function and  the X-ray
luminosity density of X-ray selected AGN at $z\approx1$.  Our analysis
provides  a lower-limit  to the  fraction  of AGN  fuelled by  secular
evolution and  minor mergers  or equivallently an  upper limit  in the
fraction  of AGN  in major  mergers.  For  this exercise  we  use deep
Chandra surveys  with available multi-waveband  Hubble Space Telescope
(HST) observations.  The  fields of choice are the  2\,Ms Chandra Deep
Field  North   (CDF-North),  the   2\,Ms  Chandra  Deep   Field  South
(CDF-South) and the  All-wavelength Extended Groth strip International
Survey  \citep[AEGIS;][]{Davis2007}.  Using  data from  these programs
has  the  advantage  of  publicly  available  follow-up  observations,
including   deep  ground-based   optical   photometry  and   extensive
spectroscopy. Throughout this paper we adopt a cosmology with $\rm H_0
= 70  \, km \,  s^{-1} \, Mpc^{-1}$,  $\rm \Omega_{M} = 0.3$  and $\rm
\Omega_{\Lambda} = 0.7$.

\section{Data}\label{sec_data}

\begin{table}
\caption{Chandra observations details}\label{tab_obs}
\scriptsize
\begin{tabular}{l p{3.cm}  c  c c c}
\hline Survey  & Obs. IDs &  Exposure & Area &  All & 0.5-7  keV\\ & &
(Ms) & ($\rm  deg^2$) & Sources & Sources  \\ (1) & (2) & (3)  & (4) &
(5) & (6) \\ \hline

CDF-N & 580, 957, 966, 967 , 1671, 2232, 2233, 2234, 2344, 2386, 2421,
2423, 3293, 3294, 3388-3391, 3408, 3409 & 1.93 & 0.11 & 516 & 273 \\

CDF-S & 581 441 582 1672  2405 2239 2312 2313 2406 2409 1431 8591-8597
9575 9578 9593 9596 9718& 1.93 & 0.06 & 428 & 245 \\

AEGIS & 3305, 4357, 4365, 5841-5854, 6210-6223, 6366, 6391, 7169 7180,
7181, 7187, 7188,  7236, 7237, 7238, 7239  & 0.19 & 0.63 &  1325 & 437
\\ \hline

\end{tabular} 
\begin{list}{}{}
\item 
The columns are: (1): Survey  name; (2): {\it Chandra} observation IDs
used  for each  survey; (3)  exposure time  in Ms  after  cleaning for
flares. In  the case of the  AEGIS survey with  multiple pointings the
median  exposure time  per pointing is  listed;  (4) total  surveyed  area in  $\rm
deg^2$;  (5) total  number of  sources in  each survey  (6)  number of
0.5-7\,keV selected  sources that overlap with the  regions covered by
the HST/ACS observations of each survey.
\end{list}
\end{table}

\subsection{X-ray observations} 

Table \ref{tab_obs} presents information on the Chandra surveys of the
AEGIS, CDF-North and CDF-South. The X-ray observations of the Extended
Groth Strip  consist of 8  ACIS-I (Advanced CCD  Imaging Spectrometer)
pointings, each with  a total integration time of  about 200\,ks split
in  at least  3 shorter  exposures obtained  at different  epochs. The
CDF-North X-ray survey consists  of 20 individual ACIS-I observations,
which sum  up to a total exposure  time of 2\,Ms. The  Chandra data in
the  CDF-South  have  recently   been  supplemented  with  new  ACIS-I
pointings, which have increased the  total exposure time in that field
to about 2\,Ms.

The Chandra observations  in the three fields above  have been reduced
and analysed in  a homogeneous way using the  methodology described by
\cite{Laird2008}.   Briefly, the  reduction used  the {\sc  ciao} data
analysis software.   After merging the individual  observations into a
single  event  file,  we  constructed  images  in  four  energy  bands
0.5-7.0\,keV  (full),  0.5-2.0\,keV  (soft), 2.0-7.0\,keV  (hard)  and
4.0-7.0\,keV  (ultra-hard).   The  count  rates in  the  above  energy
intervals are converted to fluxes in the standard bands 0.5-10, 0.5-2,
2-10 and  5-10\,keV, respectively. For  the full, hard  and ultra-hard
bands  this conversion involves  an extrapolation  from 7  to 10\,keV.
Sources in  the flux range probed  by the Chandra surveys  listed in Table
\ref{tab_obs} have, on average, spectra  similar to that of the cosmic
X-ray   Background   in   the   2-10\,keV   band,   $\rm   \Gamma=1.4$
\citep[e.g.][]{Hickox2006}.  Therefore, to  convert source count rates
to  fluxes  we  adopt  a  power-law X-ray  spectrum  with  index  $\rm
\Gamma=1.4$   absorbed  by  the   Galactic  hydrogen   column  density
appropriate for  each field. Adopting $\rm  \Gamma=1.9$ would decrease
the estimated fluxes in the 0.5-10\,keV band by about 30 per cent.

The  source detection  is a  two pass  process.  A  list  of candidate
sources is  first constructed by  running the {\sc wavdetect}  task of
CIAO  at a  low  detection threshold  ($10^{-4}$).   The total  counts
(source and background)  at the position of each  candidate source are
then extracted within the 90  per cent Encircled Energy Fraction (EEF)
radius of the Point Spread Function (PSF; see section 2.2. of Laird et
al.  2009  for details).   A local background  value is  determined by
first excluding  pixels within the 95  per cent EEF  of each candidate
{\sc wavdetect}  source and then summing  up the counts  in an annulus
centered on the source with inner radius equal to 1.5 times the 90 per
cent EEF  radius and  width of 50\,arcsec.   The probability  that the
candidate  source  is  a  random  fluctuation  of  the  background  is
estimated  assuming  Poisson   statistics.   The  final  catalogue  is
comprised of  sources with Poisson  probabilities $<4\times10^{-6}$ in
at least  one of the  energy bands defined  above.  We choose  to work
with  sources  selected  in  the  0.5-7\,keV energy  band  because  it
provides higher sensitivity resulting in  a larger sample size.  As we
will only use  X-ray sources that overlap with  the multi-waveband HST
surveys (see  below) of the  AEGIS, CDF-North and CDF-South,  in Table
\ref{tab_obs} we  list the number of 0.5-7\,keV  selected sources that
lie in the area covered by the HST observations.

\begin{figure}
\begin{center}
 \rotatebox{0}{\includegraphics[height=0.9\columnwidth]{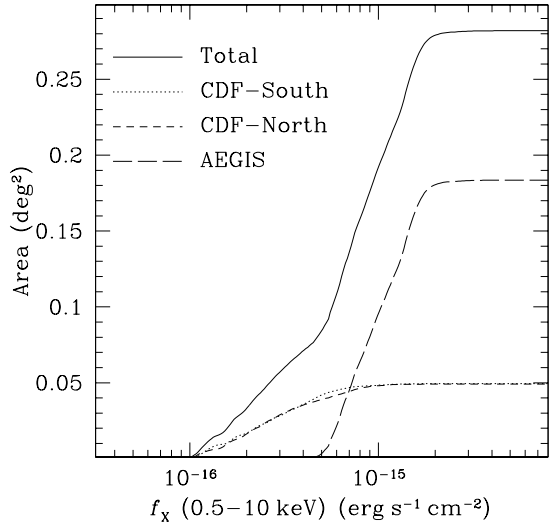}}
\end{center}
\caption{Sensitivity  curves  in  the  0.5-10\,keV band  for  the  HST
  covered   subregions   of    the   CDF-South   (dotted),   CDF-North
  (short-dashed) and  AEGIS (long-dashed). The continuous  line is the
  sum of the individual curves. }\label{fig_area}
\end{figure}

The   construction   of  the   sensitivity   maps   is  described   by
\cite{Georgakakis2008_sense}.   In  this paper  we  will  use the  1-D
representation  of the sensitivity  map, the  X-ray area  curve, which
provides an estimate  of the total survey area in  which a source with
flux  $f_X$  can be  detected.   For  simplicity  we use  area  curves
calculated in  the standard way,  i.e. by assigning a  single limiting
flux to a  detection cell, instead of the  Bayesian approach developed
by  \cite{Georgakakis2008_sense}.    The  area  curves   for  the  HST
subregions of the  AEGIS, CDF-North and CDF-South are  shown in Figure
\ref{fig_area}.

\subsection{HST photometry} 

High resolution  imaging observations from the  Hubble Space Telescope
(HST) are used to identify X-ray sources with optical counterparts and
explore the morphology of their host galaxies.

The  Advanced Camera for  Surveys (ACS)  aboard HST  has been  used to
obtain deep  images in the $V$  (F606W, 2260s) and  $I$ (F814W, 2100s)
filters over  a $\rm .1  \times 70.5 \,arcmin  ^2$ strip in  the AEGIS
field \citep{Lotz2008_aegis}.  The  5\,sigma limiting magnitudes for a
point  source are  $V_{F606W}=  28.14$ (AB)  and  $I_{F814W} =  27.52$
(AB). For extended sources these  limits are about 2\,mag brighter.  A
total of 15\,797 galaxies are detected to $I_{F814W} = 25$\,mag.

The  HST has  also surveyed  the central  most sensitive  part  of the
CDF-North and  CDF-South with the ACS  as part of  the GOODS programme
(Great Observatories Origins Deep Survey).  These observations cover a
combined  area of about  $\rm 0.08  \, deg^2$  in four  passbands, $B$
(F435W), $V$ (F606W), $i$ (F775W) and $z$ (F850LP).  The survey setup,
data    reduction    and   source    detection    is   described    by
\cite{Giavalisco2004}.  We  use the version  v1.0 of the  mosaiced ACS
images and the version v1.1 of the ACS multi-band source catalogs. The
total exposure  time of this data  release is about  7200, 5000, 5000,
and 10660\,s  in the $B$, $V$,  $i$ and $z$  filters respectively.  We
use the  $z$-band selected catalogues  which include a total  of 32048
and 29599 sources in the CDF-North and the CDF-South respectively.

The Likelihood Ratio method \citep[LR;][]{Sutherland_and_Saunders1992,
  Ciliegi2003, Laird2008}  is adopted  to identify X-ray  sources with
optical counterparts.  In this  exercise we estimate the ratio between
the  probability that a  source, at  a given  distance from  the X-ray
position and with  a given optical magnitude, is  the true counterpart
and the probability that the source is a spurious alignment.

\begin{equation}\label{eq_lr} {\rm LR}=\frac{q(m)\,f(r)}{n(m)},
\end{equation}

\noindent where  $q(m)$ is the expected magnitude  distribution of the
true  optical  counterparts, $f(r)$  is  the probability  distribution
function of  the positional  uncertainties in both  the X-ray  and the
optical  source catalogues  and $n(m)$  is the  background  density of
optical galaxies of magnitude $m$.

For the positional  accuracy of the X-ray sources  we adopt a Gaussian
distribution  with  standard  deviation  estimated as  a  function  of
off-axis   angle  and  total   number  of   counts  as   described  by
\cite{Laird2008}. The  positional uncertainty of the  HST source lists
is  also assumed  to follow  the normal  distribution with  FWHM (Full
Width Half  Maximum) of 0.07\,arcsec  for AEGIS \citep{Lotz2008_aegis}
and 0.12\,arcsec for the CDF-North and South. In equation \ref{eq_lr},
$f(r)$ is therefore  assumed to be a Gaussian  with standard deviation
equal to the combined optical and X-ray source positional errors added
in quadrature.

The a priori probability $q(m)$ that an X-ray source has a counterpart
with magnitude $m$, is determined  as follows.  Firstly, the X-ray and
optical source  calalogues are matched  by simply finding  the closest
counterpart within  a fixed search  radius of 2\,arcsec.   Secondly, a
total of 100  mock catalogs are constructed by  randomising the source
positions of the  HST catalog and the cross-matching  is repeated. The
magnitude distribution  of the counterparts  in the mock and  the real
catalogs are subtracted to determine the magnitude distribution of the
true associations, $q(m)$.

Having  estimated   $f(r)$  and   $q(m)$  we  identify   all  possible
counterparts to X-ray sources within  a 4\,arcsec radius and the LR of
each  one is  estimated using  equation \ref{eq_lr}.   We  consider as
counterparts those galaxies with LR above a certain limit.  The choice
of  the  cutoff  in LR  is  a  trade  off  between maximum  number  of
counterparts  and minimum  spurious identification  rate. In  order to
assess how  secure an  optical counterpart is  we use  the reliability
parameter defined by \cite{Sutherland_and_Saunders1992}

\begin{equation}\label{eq_rel}   \rm  Rel_i=\frac{LR_i}{\sum_{j}  LR_j
+(1-Q)},
\end{equation}

\noindent  where $\rm  Rel_i$ is  the reliability  of the  $i$ optical
counterpart of  an X-ray source, the  index $j$ of  the summation runs
over all the  possible counterparts within the search  radius and Q is
the fraction  of X-ray sources  with identifications to  the magnitude
limit  of the optical  survey. It  can be  shown that  the sum  of the
reliabilities of individual counterparts equals the expected number of
true associations  (Sutherland \&  Saunders 1992). Comparison  of $\rm
\sum_i  Rel_i$  with  the  total  number  of  counterparts  with  $\rm
LR>LR_{limit}$  provides an  estimate of  the  spurious identification
rate.   By varying  $\rm LR_{limit}$  one can  minimise the  number of
false associations.   We chose to  use $\rm LR_{limit}=0.5$.   At this
cutoff, 78  (340/437), 74 (201/273) and  81 (199/245) per  cent of the
0.5-7\,keV  selected  X-ray  sources   in  the  AEGIS,  CDF-North  and
CDF-South,  respectively,  have  counterparts.   These  fractions  are
estimated using  the total number  of X-ray sources that  overlap with
the   HST   surveys  of   these   fields.    The  estimated   spurious
identification rate is $<5$ per cent.

\subsection{Optical spectroscopy}

Optical spectroscopy of X-ray sources  in the AEGIS field is primarily
from  the  DEEP2 redshift  survey  \citep{Davis2003},  which uses  the
DEIMOS spectrograph  \citep{Faber2003} on the  10\,m Keck-II telescope
to  obtain  redshifts  for  galaxies  to $R_{AB}  =  24.1$\,mag.   The
observational  setup   uses  a  moderately   high  resolution  grating
($R\approx5000$),  which   provides  a  velocity   precision  of  $\rm
30\,km\,s^{-1}$  and a wavelength  coverage of  6500--9100\,\AA.  This
spectral  window  allows  the  identification of  the  strong  [O\,II]
doublet  3727\AA\, emission line  to $z<1.4$.   We use  DEEP2 galaxies
with redshift  determinations secure  at the $>90\%$  confidence level
\citep[quality   flag   $Q   \ge   3$;][]{Davis2007}.    Spectroscopic
observations targeting  X-ray sources in AEGIS have  also been carried
out at the  MMT using the Hectospec fiber  spectrograph.  Full details
about these observations are given by \cite{Coil2009}.  A total of 449
X-ray sources were targeted  with 5 Hectospec configurations. The data
were taken in queue mode in  May 2007, July 2007, and May 2008.  Total
integration  times were  2  hours per  configuration.  The  wavelength
coverage is  $\sim4500-9000$ \AA  \ at 6  \AA \ resolution.   The data
were  reduced  using  the  HSRED IDL  reduction  pipeline\footnote{See
  http://www.astro.princeton.edu/$\sim$rcool/hsred}.   Redshifts  were
measured either from emission lines from the AGN itself or HII regions
if the  host galaxy  has ongoing  star formation, or  from the  Ca H+K
absorption features  for lower  luminosity AGN in  early-type galaxies
with little or  no star formation.  Although objects  were targeted to
$R_{AB}=25$, the  secure redshift identification rate was  50 per cent
at  $R_{AB}=22.5$\,mag  and 15  per  cent  at $R_{AB}=24.5$\,mag.   We
obtained  high-confidence redshifts for  230 sources,  or 51\%  of the
targeted   sample.   The   resulting  redshifts   lie  in   the  range
$0<z<4$. The  original Groth  Strip has been  targeted by a  number of
spectroscopic  programs  \citep{Lilly1995,Brinchmann1998, Hopkins2000,
  Vogt2005}  that  have  been  compiled  into  a  single  database  by
\cite{Weiner2005}\footnote{http://saci.ucolick.org/verdi/public/index.html}.
The entire AEGIS overlaps with the Sloan Digital Sky Survey (SDSS) and
therefore  spectra for relatively  bright galaxies  and QSOs  are also
available  \citep{York2000}.  All the  above diverse  datasets provide
high  confidence spectroscopic redshifts  for 159  0.5-7\,keV selected
X-ray sources in the HST/ACS subregion of the AEGIS.

Optical spectroscopy in the CDF-N  and in the GOODS-North is available
from either  programmes that specifically target  the X-ray population
in these  fields \citep[e.g.][]{Barger2003, Barger2005,  Cowie2003} or
the  Keck Treasury  Redshift  Survey \citep[TKRS;][]{Wirth2004}.   The
TKRS  uses the  DEIMOS spectrograph  \citep{Faber2003} at  the Keck-II
telescope to  observe optically  selected galaxies to  $R_{AB} \approx
\rm 24.4\,  mag$ in the GOODS-North. The  publicly available catalogue
consists of about 1400 secure  redshifts. Out of the 201 X-ray sources
selected  in  0.5-7\,keV  band  and  optical  counterparts  with  $\rm
LH>0.5$, a total of 149 have spectroscopic redshifts.

There  are  many  spectroscopic  campaigns in  the  CDF-South.   These
include follow-up observations targeting specific populations, such as
X-ray   sources   \citep{Szokoly2004},   $K$-band  selected   galaxies
\citep{Mignoli2005},      and       high      redshift      candidates
\citep[e.g.][]{Dickinson2004, Stanway2004a,  Stanway2004b}, as well as
generic  spectroscopic surveys  of faint  galaxies in  the  GOODS area
using   the  FORS2   \citep{Vanzella2005,   Vanzella2008}  and   VIMOS
\citep{LeFevre2004, Popesso2008} spectrographs  at the VLT.  There are
151 spectroscopic  redshift estimates  of the 199  0.5-7\,keV selected
X-ray sources with HST optical identifications.

\subsection{Ground-based imaging and photometric redshifts}

The  spectroscopic   observations  are  complemented   by  photometric
redshifts calculated  using ground-based imaging of  the fields listed
in Table  \ref{tab_obs}. This  provides a redshift  (spectroscopic and
photometric) completeness  of 100 per  cent for the  magnitude limited
sub-sample used in this paper  (see next section for details on sample
selection).  We prefer the  ground-based observations over the HST/ACS
data for  the photometric redshift estimation because  they are deeper
(AEGIS,  CDF-North)  and/or  have  more photometric  bands,  including
imaging in the $U$ filter.

The   AEGIS   is   one   of   the  fields   of   the   deep   synoptic
Canada-France-Hawaii   Telescope   Legacy   Survey  (CFHTLS).    These
observations cover $\rm 1\,deg^2$ in  the Extended Groth Strip in five
filters   ($ugriz$).   We  use   the  CFHTLS   data  from   the  T0003
release\footnote{http://terapix.iap.fr/article.php?idarticle=556.},
which rearches  a limiting magnitude  of $\rm i_{AB}\approx26.5\,mag$.
We use  the photometric  redshifts estimated by  \cite{Ilbert2006} for
sources brighter than $i_{AB}\approx25$\,mag.  The CFHTLS data overlap
fully  with area surveyed  by the  HST, thereby  providing photometric
redshifts for all HST sources with $\rm i_{AB}\la 25\,mag$.

Deep  multiwavelength imaging  ($UBVRIz$)  in the  CDF-North has  been
presented by Capak et al.  (2004). These observations cover about $\rm
0.2\, deg^2$ and  extend beyond the CDF-North field  of view.  In this
study we  use the $R$-band selected sample  comprising 47\,451 sources
to  the limit  $R_{AB}=26.6$\,mag  ($5\sigma$). Photometric  redshifts
using these  data have been presented  by \cite{Georgakakis2007} using
the the methods described by \cite{rowan2005, rowan2008}.

Ground-based  photometric observations  in  17 narrow  and broad  band
filters have  been obtained in an  area of $\rm  1\,deg^2$ centered on
the  CDF-South  as  part  of  the  COMBO-17  survey  \citep{Wolf2004}.
Although these observations are not  as deep as the HST photometry the
large  number  of  filters  provides  excellent  photometric  redshift
estimates to $R\approx24$\,mag \citep{Wolf2004}.

\begin{figure}
\begin{center}
 \rotatebox{0}{\includegraphics[height=0.9\columnwidth]{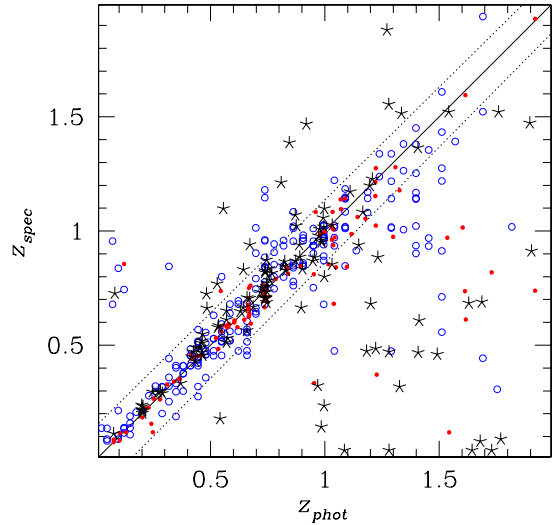}}
\end{center}
\caption{Photometric against spectroscopic  redshift estimates for the
  X-ray  sources in  the CDF-South  (red dots),  CDF-North  (open blue
  circles)  and  AEGIS (black  stars).   The  continuous  line is  the
  $z_{phot}=z_{spec}$ relation,  while the dashed  lines correspond to
  $\delta  z=\pm0.2$.  All  sources with  spectroscopic  redshifts are
  plotted without applying any magnitude limit.  }\label{fig_photoz}
\end{figure}

Figure   \ref{fig_photoz}  compares  photometric   with  spectroscopic
redshift estimates  for X-ray sources in the  CDF-South, CDF-North and
AEGIS.     Catastrophic    redshifts,    defined   as    those    with
$(z_{spec}-z_{phot})/(1+z_{spec})>0.15$, represent  13, 16 and  32 per
cent of  the photometric redshift  determinations of X-ray  sources in
the CDF-South, CDF-North and  AEGIS, respectively. The accuracy of the
photometric redshifts is  estimated by the rms values  of the quantity
$(z_{spec}-z_{phot})/(1+z_{spec})$    after   excluding   catastrophic
redshifts.   This is estimated  0.04, 0.06  and 0.05  for each  of the
surveys above, respectively. The CFHTLS photometric redshifts of X-ray
sources suffer  the largest catastrophic  redshift rate.  This  is not
surprising  as  these  redshifts  have  been  estimated  using  galaxy
templates  only \citep{Ilbert2006} and  also use  a smaller  number of
photometric  bands compared  to  the CDF-North  and South  photometric
redshifts.

Uncertainties in the photometric redshifts are expected to have only a
small impact on the results. As  described in the next section, in the
analysis  we  use  a  magnitude-limited sub-sample  of  X-ray  sources
($I_{AB}<23$\,mag  for the  AEGIS and  $z_{AB}<24$\,mag for  the GOODS
fields), for which the spectroscopic redshift completeness is about 88
per cent.

\section{Sample selection and morphological classification}\label{sec_sample}

The sample consists  of X-ray sources selected in  the 0.5-7\,keV band
and   in  the   redshift  (photometric   or   spectroscopic)  interval
$z=0.5-1.3$.  The  median redshift of the sample  is $\approx0.8$ (see
below), thereby allowing study of AGN hosts close to the peak of X-ray
luminosity density of the Universe (e.g. Ueda et al. 2003).

Two main approaches are adopted  in the literature for quantifying the
optical     morphology     of     galaxies.      Visual     inspection
\citep[e.g.][]{Bell2005}  and automated classification  schemes, which
measure  morphology-sensitive  structural   parameters  of  the  light
distribution   of   galaxies.   The   latter   methods  include   both
non-parametric morphology estimators \citep{Abraham1996, Lotz2004} and
model fits to the optical profile of galaxies \citep[e.g.][]{Peng2002,
  Simard2002}.  These techniques have the advantage of objectivity and
repeatability   of  the   measurements.   However,   there   are  also
limitations.  In the case of AGN for example, the nuclear point source
can  be  sufficiently bright  to  bias  the  estimation of  structural
parameters  of   the  host  galaxy  light  profile.    The  AGN  light
distribution (usually a point source) has to be modeled accurately and
then  subtracted from  the image,  before the  morphology of  the host
galaxy is  determined.  Although there  has been progress  recently in
this direction  \citep[e.g.][]{Kuhlbrodt2004, Simmons2008, Gabor2008},
the decomposition of the nuclear  point source from the host galaxy is
not  straightforward, especially  in the  case of  real  galaxies with
complex profiles, including  disks, bulges, star-forming regions, dust
lanes,   etc.   Additionally,  non-parametric   morphology  estimators
typically produce  broad morphological classes.  Irregular/interacting
systems for example, maybe be hard to discriminate from late-type disk
galaxies with bright star-forming regions superimposed on their spiral
arms. It  is also  unclear how methods  that fit models  (e.g.  Sersic
function) to  the galaxy light profile treat  irregular or interacting
galaxies and whether they can isolate such systems.

Eyeballing has issues of  its own \citep[e.g.][]{Lotz2008_sim}, but is
best  suited  to  this  study which  requires  discrimination  between
spirals, bulges and  irregular/interacting systems. Also, the presence
of a photometrically important bulge  or nuclear point source may bias
automated morphological classification  schemes against disks, even if
they are large and obvious to  the eye. We therefore choose to proceed
with visual inspection of the HST images of X-ray AGN to classify them
into  4  groups,  disks  (i.e.   spirals),  early-types  (i.e.   bulge
dominated), peculiar/interacting and  point sources.  The latter class
includes systems where  the central AGN outshines the  host galaxy and
therefore morphological classification is  not possible.  This type of
sources pose a challenge for any method that attempts to determine the
light   distribution  of   the  underlying   AGN  host   galaxy.   The
morphological  classes  have been  determined  independently by  three
classifiers (AG,  ESL, ALC).   In the analysis  that follows  we apply
equal  weights to each  of the  three independent  classifications. We
caution that faint or  red disk-dominated sources may be misclassified
as  early-type hosts,  thereby introducing  a systematic  bias against
disks in our visual morphological classification.  However, comparison
with the classifications based on  the Sersic index suggests that this
bias is not affecting the results and conlcusions (see section 5).

\begin{figure*}
\begin{center}
 \rotatebox{0}{\includegraphics[height=1.9\columnwidth]{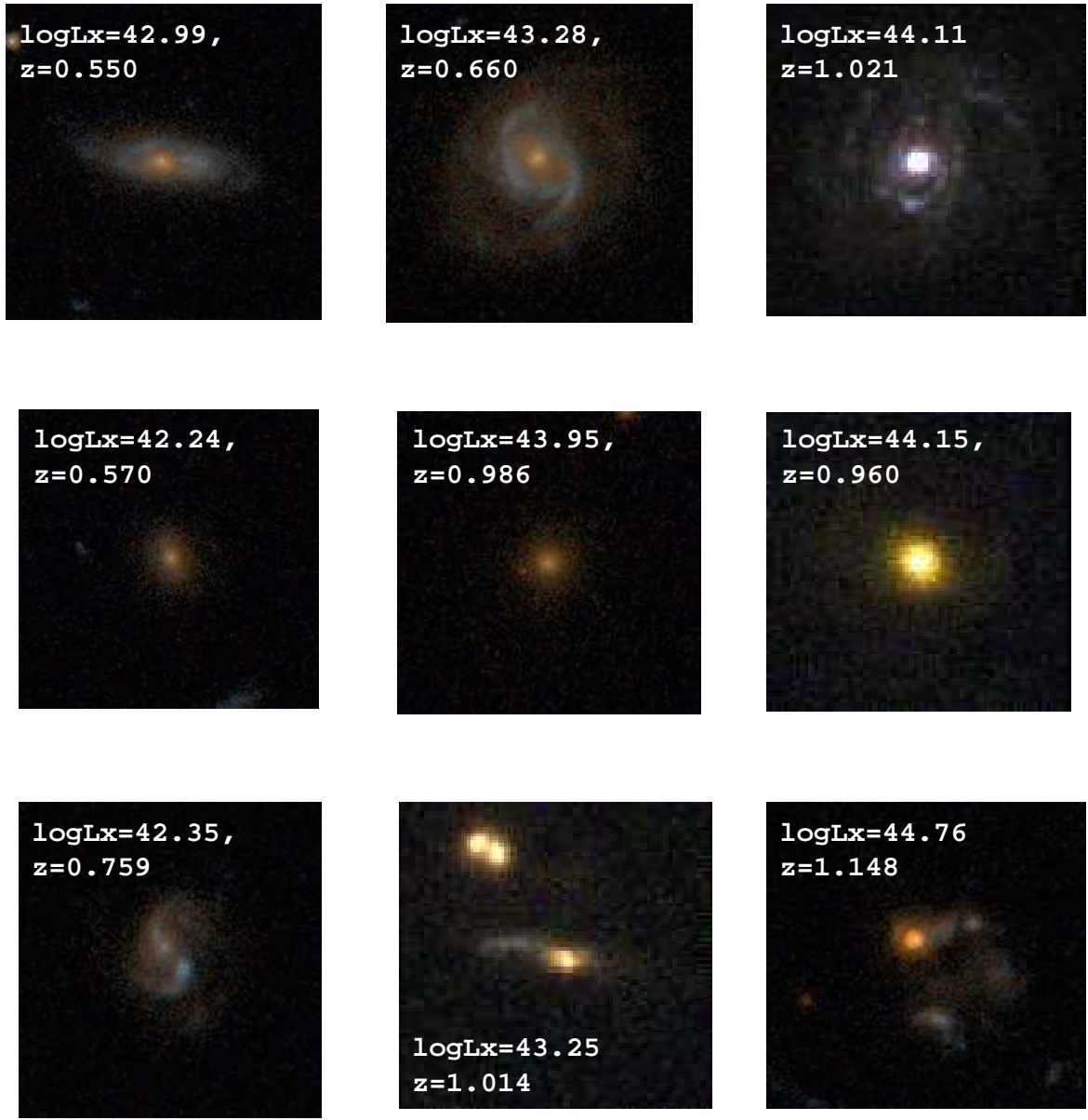}}
\end{center}
\caption{Examples of the  HST true colour images of  AGN classified as
  disks  (top  row),   early-type  (middle)  and  peculiar/interacting
  (bottom). The  X-ray luminosity  (0.5-10\,keV) and redshift  of each
  source  are also  shown.   The images  are about  $5\times5$\,arcsec
  in size. }\label{fig_morph_examples}
\end{figure*}

We have empirically determined that reliable morphological classes are
possible to  $I_{AB}=23$\,mag for  the AEGIS and  $z_{AB}=24$\,mag for
the deeper  GOODS HST observations.   Therefore, in the  analysis that
follows  we use  a  magnitude-limited subsample  of  454 (CDF-N:  156;
CDF-S: 130;  AEGIS: 168) X-ray  sources brighter than  $I_{AB}=23$ and
$z_{AB}=24$\,mag  for the  AEGIS  and the  GOODS fields  respectively.
This choice of  magnitude limits has the additional  advantage of high
spectroscopic  redshift completeness,  thereby minimising  the  use of
photometric  redshifts. Secure  spectroscopic redshifts  are available
for 398  of the 454  sources, of which  7 are Galactic stars.   We use
photometric  redshifts for  56  sources only  (CDF-N:  14; CDF-S:  13;
AEGIS:  29)   and  therefore   uncertainties  in  the   estimation  of
photometric  redshifts have  a small  impact  on the  results.  If  we
further constrain the sample  to the redshift interval $z=0.5-1.3$ the
total number of X-ray sources  is 266 (median redshift 0.8).  About 30
per cent  of the sources in  that sample have  been assigned different
morphological  classifications by  the three  independent classifiers.
We have  tested that these differences  do not affect  the results and
conclusions of the  paper. In the next section  for example, the X-ray
luminosity function  for different host galaxy  morphological types is
estimated.  It is found that  the X-ray luminosity functions for the 3
independent morphological  classifications agree within  the $1\sigma$
Poisson  uncertainty.  Examples  of the  HST images  of  X-ray sources
classified as  spirals, bulges and peculiars/interacting  are shown in
Figure \ref{fig_morph_examples}.

\begin{figure}
\begin{center}
\rotatebox{0}{\includegraphics[height=0.9\columnwidth]{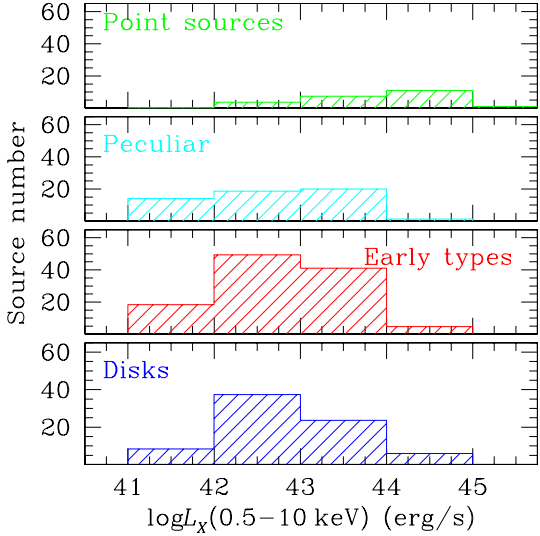}}
\end{center}
\caption{The X-ray  luminosity distribution  of X-ray selected  AGN in
  the  redshift interval  $0.5<z<1.3$  as a  function  of host  galaxy
  morphology.}\label{fig_morph_hist}
\end{figure}


\section{Luminosity Function}\label{sec_xlf}

The  binned X-ray luminosity  function is  derived using  the standard
non-parametric  $\rm 1/V_{max}$  method  \citep{Schmidt1968}. In  this
calculation  we  take  into  account  the  X-ray  selection  function,
parameterised by the X-ray area curve, and the optical magnitude limit
for morphological classification.  The luminosity function, $\phi$, in
logarithmic luminosity bins, $d\log L$, is estimated from the relation

\begin{equation} \phi \, d\log L = \sum_{i} \frac{1}{V_{max,i}
(L_X,M,z)},
\end{equation}

\noindent  where $V_{max,i}(L_X,M,z)$ is  the maximum  comoving volume
for which the source $i$ with X-ray luminosity $L_X$, absolute optical
magnitude  $M$  and  redshift  $z$,  satisfies  the  sample  selection
criteria, i.e.  redshift range,  apparent optical magnitude cutoff for
morphological  classification  and  X-ray  flux  limit.   The  maximum
comoving volume is estimated from the relation

\begin{equation} V_{max,i}(L_X,M,z) = \frac{c}{H_0} \int_{z1}^{z2} \,
\Omega(L_X,z)\, \frac{dV}{dz}\,dz\, dL,
\end{equation}

\noindent where  $dV/dz$ is the  volume element per  redshift interval
$dz$.   The integration limits  are $z1=z_L$  and $z2=min(z_{optical},
z_U)$, where we have defined $z_L$, $z_U$ the lower and upper redshift
limits of  the sample and $z_{optical}$  is the redshift  at which the
source  will become  fainter  than the  optical  magnitude cutoff  for
morphological  classification. $\Omega(L_X,z)$ is  the solid  angle of
the  X-ray survey available  to a  source with  luminosity $L_X$  at a
redshift  $z$  (corresponding  to  a  flux $f_X$  in  the  X-ray  area
curve). The uncertainty of a given luminosity bin is

\begin{equation} \delta \phi^2 = \sum_{i} \left ( \frac{1}{V_{max,i}
(L_X,M,z)} \right )^2.
\end{equation}

\noindent In addition to  the Poisson uncertainties above, differences
among the  morphological classes  determined by the  three independent
classifiers are  also accounted for  in the error budget.   We perform
1000 experiments in  which the XLF is determined  by randomly choosing
for  each   source  one  of  the  three   morphological  types.   This
calculation  provides  for each  luminosity  bin  an  estimate of  the
$1\sigma$ rms of the XLF because of uncertainties in the morphological
classifications.   This is  then  added quadratically  to the  Poisson
errors above. We note  however, that Poisson errors typically dominate
the uncertainties.

In order  to convert  absolute to apparent  optical magnitude  for the
1/Vmax  calculaton,  optical k-corrections  for  both  galaxy and  QSO
templates   are   used   depending   on   the   source   morphological
classification.   For the early,  disk and  peculiar classes  we adopt
respectively  the   E/S0,  Scd  and   Im  galaxy  type   templates  of
\cite{Coleman1980}, while for point  sources we use the SDSS composite
QSO spectrum of \cite{VandenBerk2001}.   For AEGIS sources the optical
k-corrections are estimated for  the F814W filter, while for CDF-North
and South X-ray AGN the F850LP filter is used in the calculations.

The  X-ray luminosities  in  the 0.5-10\,keV  band  are corrected  for
obscuration.   The hardness  ratio,  HR, estimated  in  the 0.5-2  and
2-7\,keV bands,  is used  to determine the  column density,  $N_H$, of
individual sources assuming an intrinsic power-law X-ray spectrum with
index $\Gamma=1.9$ \citep[e.g.][]{Nandra1994}.  These column densities
are  then used  to  correct  the observed  $L_X$  for absorption.   We
caution that for unobscured  sources the assumption of $\Gamma=1.4$ in
section 2 overstimates the fluxes  and hence the luminosities by about
30 per cent  or $\rm \Delta \log L_X=0.1$.   This effect therefore has
only  a  minor impact  on  the  estimated  $L_X(\rm 0.5-10  \,  keV)$.
Additionally,  unobscured sources ($\rm  N_H<10^{22} \,  cm^{-2}$) are
only a  small fraction of  the X-ray population  at the depths  of the
CDF-North/South  \citep[$\approx  20\%$;][]{Akylas2006}  and the  AEGIS
\citep[$\approx  40\%$;][]{Georgakakis2006}. When estimating  the X-ray
k-correction of individual sources for the $\rm 1/V_{max}$ calculation
we also  take into account the  intrinsic source $N_H$  by adopting an
absorbed power-law spectral  energy distribution with $\Gamma=1.9$ and
photoelectric    absorption   cross    sections   as    described   by
\cite{Morrison1983} for solar metallicity.

\section{Results}\label{sec_results}

Figure \ref{fig_morph_hist} plots the X-ray luminosity distribution of
different  AGN host  galaxy  morphological types.   In agreement  with
previous studies, early-type  bulge-dominated galaxies are the largest
group, contributing  $42\pm4$ per cent of the  overall AGN population.
These systems are either major merger remnants observed after the peak
of  the  activity \citep[e.g.][]{Hopkins2006}  or  galaxies that  have
recently  accreted cold  gas,  through minor  interactions, which  has
resulted   in  the  re-activation   of  the   SMBH  at   their  centres
\citep[e.g.][]{Khalatyan2008}.  Alternative models for the fuelling of
the SMBH in  early type galaxies inlcude recycled  gas from dying stars
\citep[e.g.][]{Ciotti_Ostriker2007}  or accretion  from a  cooling hot
halo         \citep[e.g.][]{Croton2006,        Cattaneo_Teyssier2007}.
\cite{Georgakakis2008_stack} have  shown that the  Asymmetry parameter
\citep{Abraham1996} distribution  of X-ray AGN is  broader compared to
optically   selected  early-type   galaxies,   suggesting  low   level
morphological disturbances. This finding favors the scenario where AGN
in early type  hosts are associated with past  merger events (minor or
major).

Isolated disk galaxies represent about $28\pm3$ per cent of all AGN in
Figure \ref{fig_morph_hist}.  These are systems in which the accretion
onto the SMBH is likely to  be triggered either by secular processes or
minor tidal disruptions.  Peculiars are about $20\pm3$ per cent of the
AGN in our sample.  These systems include ongoing major merger events.
Their fraction may  indicate the time-scale that major  mergers can be
identified    morphologically   \citep[e.g.     $\rm    \la   10^{9}\,
  yrs$;][]{Lotz2008_sim}.   We  caution  however, that  this  fraction
provides  only  a upper  limit  to this  time-scale  as  the group  of
peculiars may also include Irregular galaxies that are not necessarily
the product  of major mergers.   Finally, point-like sources  in which
the light from  the central engine outshines the  host galaxy and does
not allow  morphological classification are a minority  in our sample,
$8\pm2$  per cent.   As expected  these sources  are more  numerous at
bright X-ray luminosities.   We note that the errors  in the fractions
above  include uncertainties  in the  morphological  classification of
individual sources  (see previous section) in addition  to the Poisson
errors.

\begin{figure}
\begin{center}
\rotatebox{0}{\includegraphics[height=0.9\columnwidth]{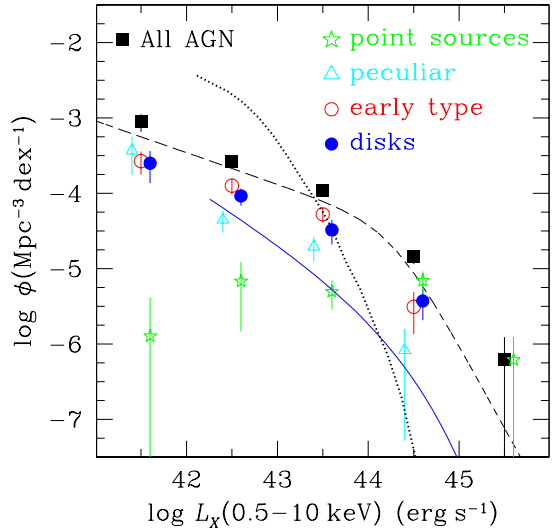}}
\end{center}
\caption{The 0.5-10\,keV X-ray  luminosity function for different host
  galaxy  types.   Stars (green)  correspond  to  point sources,  open
  triangles (cyan)  are peculiar systems,  open circles (red)  are for
  early-type  hosts  and  filled  circles (blue)  correspond  to  disk
  galaxies.   For   clarity  triangles  are   offset  horizontally  by
  --0.1\,dex in $\log L_X$, while stars and  filled circles are offset by +0.1\,dex.
  The  filled squares  (black)  represent the  total X-ray  luminosity
  function, independent of host galaxy type. This is compared with the
  dashed line,  which is the Barger  et al.(2005) XLF  at $z=0.8$ (the
  median redshift  of the sample)  estimated in the 2-8\,keV  band and
  converted  to  the  0.5-10\,keV  band  assuming  a  power-law  X-ray
  spectrum with  photon index $\Gamma=1.9$.   The dotted curve  is the
  ``maximal  evolution'' XLF  prediction of  the  stochastic accretion
  mode of Hopkins \& Hernquist (2006) at $z=1$.  This curve is adapted
  from Figure 6 of Hopkins \& Hernquist (2006) adopting the Marconi et
  al.  (2004)  bolometric corrections.  The continuous  (blue) line is
  the updated version of the XLF for the stochastic accretion mode (P.
  F.   Hopkins priv.   communication)  estimated using  a revised  AGN
  lightcurve,  which  results  in  more  rapid evolution  of  the  AGN
  emission  after  the  peak  of  the accretion  (see  text  for  more
  details).}\label{fig_xlf}
\end{figure}

\begin{figure}
\begin{center}
\rotatebox{0}{\includegraphics[height=0.9\columnwidth]{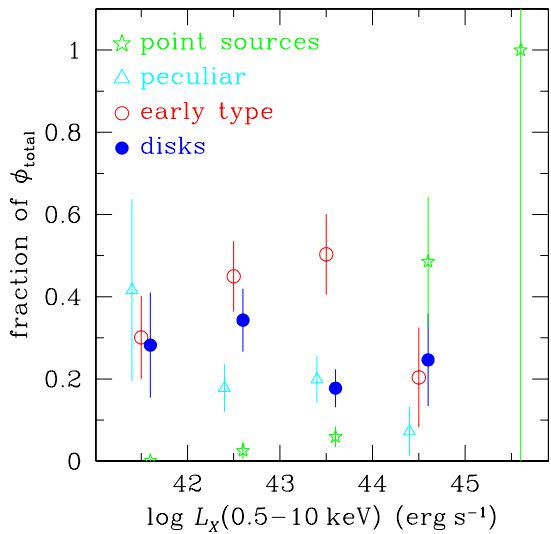}}
\end{center}
\caption{The 0.5-10\,keV X-ray  luminosity function for different host
  galaxy  types  normalised  to  the  total XLF.   The  vertical  axis
  measures the contribution of different galaxy types to the XLF.  The
  symbols are the same as in Figure \ref{fig_xlf}.}\label{fig_rel_xlf}
\end{figure}

The  XLF, split  into  different host  galaxy  morphological types  is
plotted   in  Figure   \ref{fig_xlf}   and  is   presented  in   Table
\ref{tab_xlf}. The XLF  of all AGN in the sample  is in good agreement
with   the   recent   determination  of   \cite{Barger2005}.    Figure
\ref{fig_rel_xlf}  and  Table  \ref{tab_xlf1} present  the  fractional
contribution  of  different  morphological  types to  the  total  XLF.
Early-type  hosts are a  major component  of the  XLF over  almost the
entire  luminosity  range,  while  point sources  become  dominant  at
$L_X(\rm  0.5-10\,keV)>10^{44}  \rm \,  erg  \,  s^{-1}$.  Disk  hosts
represent between 25-30\% of the  total XLF.  The fraction of Peculiar
hosts to the  total XLF shows a weak  decreasing trend with increasing
luminosity, which is, however, within the uncertainties of the data.

\begin{figure}
\begin{center}
\rotatebox{0}{\includegraphics[height=0.9\columnwidth]{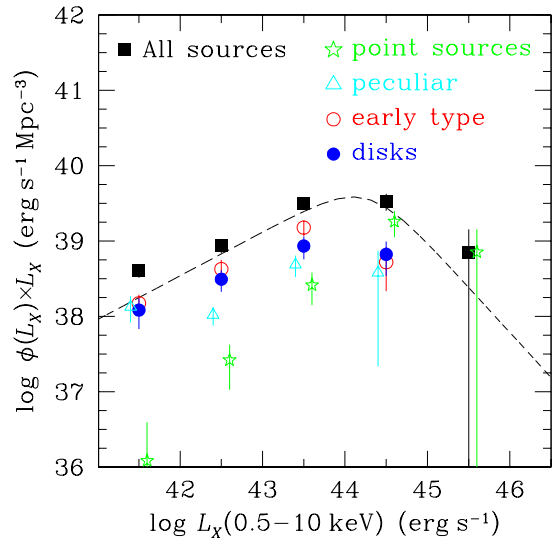}}
\end{center}
\caption{The 0.5-10\,keV  X-ray luminosity density  for different host
  galaxy  types. The  symbols and  curves are  the same  as  in Figure
  \ref{fig_xlf}.}\label{fig_em}
\end{figure}

\begin{figure}
\begin{center}
\rotatebox{0}{\includegraphics[height=0.9\columnwidth]{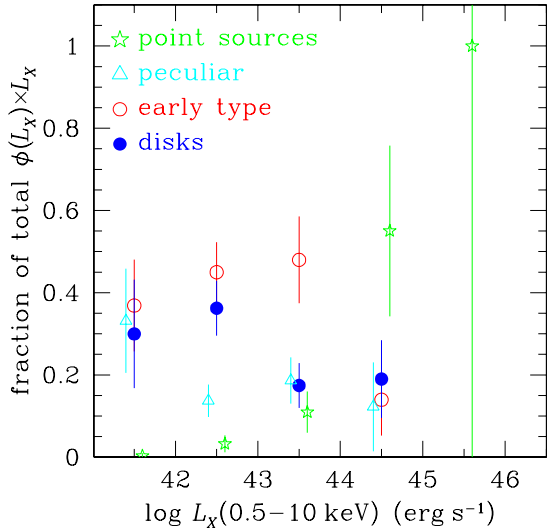}}
\end{center}
\caption{The  fractional  contribution of  different  AGN host  galaxy
  morphological types to the 0.5-10\,keV X-ray luminosity density. The
  symbols are the same as in Figure \ref{fig_xlf}.}\label{fig_rel_em}
\end{figure}

The contribution  of different  morphological types to  the luminosity
density,  $\phi  \times  L_X$,  at  $z\approx1$ is  shown  in  Figures
\ref{fig_em} and  \ref{fig_rel_em}. The results are  also presented in
Tables  \ref{tab_xlf} and  \ref{tab_xlf1}.  Early-type  hosts dominate
the luminosity  density up to $L_X(\rm 0.5  - 10 \, keV)  = 10^{44} \,
erg  \, s^{-1}$.   Above this  limit point-like  sources,  i.e.  QSOs,
become the dominant component of $\phi \times L_X$.  Disk and peculiar
host  galaxies  represent  about  11-35  per cent  of  the  total  AGN
luminosity density.

We have  checked how  sensitive the results  above are to  the adopted
method for determining  the morphology of the AGN  host galaxies.  The
classifications  based on  visual inspection  are compared  with those
determined by fitting a single  Sersic function to the AGN host galaxy
light distribution  (Griffith et  al. in prep)  using the  Galfit code
\citep{Peng2002}. A cutoff in the Sersic index, $n=2.5$, is applied to
discriminate  between  disk ($n<2.5$)  and  bulge ($n>2.5$)  dominated
galaxies.  The  Sersic model fits  are reliable for systems  where the
central  AGN does  not  dominate  over the  host  galaxy light.   This
requirement translates to an empirically determined (visual inspection
of the HST images) X-ray luminosity cut  of $L_X (\rm 0.5 - 10 \, keV)
\approx 10^{44} \, erg \,  s^{-1}$.  The comparison between the visual
and Sersic  index classifications  is therefore restricted  to sources
with $L_X (\rm 0.5 - 10 \,  keV) < 10^{44} \, erg \, s^{-1}$.  We only
consider sources  that are visually classified  ``early'' or ``disk''.
The class of  ``peculiars'' is excluded because they  comprise a large
fraction of disturbed interacting/merging systems with complex optical
light  distributions.  ``Point''  sources are  also excluded  to avoid
uncertainties in the Sersic function fits associated with the presence
of  bright  emission from  the  nuclear  galaxy  regions.  The  Sersic
indeces  of  both ``peculiars''  and  ``point  sources''  are hard  to
interpret in the context of disk or bulge dominated light profiles. It
is  found  that  although  there  is a  discrepancy  between  the  two
classification methods (eyeballing vs  Sersic index) for about 30\% of
the bulge and disk-dominated sources  in the sample, the main results,
e.g.  the XLF of disks and bulges, are consistent within the $1\sigma$
errors and  do not  depend on the  adopted method for  determining the
morphology of AGN hosts.

\nocite{Marconi2004} \nocite{Barger2005}

\begin{table*} 
\caption{AGN XLF and luminosity density as a function of morphological
  type}\label{tab_xlf}
\scriptsize
\begin{tabular}{c cccc  cccc}
\hline

$\log      L_X$      &\multicolumn{4}{c}{$\log      \phi(L_X)$}      &
\multicolumn{4}{c}{$\log ( L_X\,\phi(L_X)  )$}\\ ($\rm erg\,s^{-1}$) &
\multicolumn{4}{c}{($\rm    \times    10^{-6}    \,   Mpc^{-3}$)}    &
\multicolumn{4}{c}{($\rm  \times  10^{37} \,  erg\,s^{-1}\,Mpc^{-3}$)}
\\ & disk & early & peculiar & point & disk & early & peculiar & point
\\ \hline

41--42   &  $-3.60\pm0.19$   &  $-3.57\pm0.14$   &   $-3.43\pm0.23$  &
$-5.89\pm0.97$  &  $1.08\pm0.19$  &  $1.17\pm0.13$ &  $1.12\pm0.16$  &
$-0.91\pm0.97$   \\  42--43  &   $-4.03\pm0.11$  &   $-3.89\pm0.10$  &
$-4.35\pm0.13$  &  $-5.16\pm0.34$ &  $1.49\pm0.13$  & $1.62\pm0.13$  &
$1.01\pm0.2$   &    $0.42\pm0.25$\\   43--44   &    $-4.48\pm0.15$   &
$-4.27\pm0.09$  & $-4.70\pm0.15$  & $-5.30\pm0.18$  &  $1.93\pm0.14$ &
$2.17\pm0.10$   &   $1.68\pm0.13$   &   $1.41\pm0.19$  \\   44--45   &
$-5.42\pm0.19$  & $-5.50\pm0.24$ &  $-6.08\pm0.40$ &  $-5.15\pm0.14$ &
$1.82\pm0.20$   &  $1.72\pm0.25$   &  $1.58\pm0.40$   &  $2.26\pm0.16$
\\  45--46  & --  &  --  & --  &  $-6.20\pm0.43$  & --  &  --  & --  &
$1.85\pm0.43$ \\

\hline
\end{tabular} 
\begin{list}{}{}
\item 
The columns are: (1): X-ray luminosity interval; (2): logarithm of XLF
in units  of $\rm \times  10^{-6} \, Mpc^{-3}$ for  disks, early-type,
peculiar and  point sources; (3)  luminosity density in units  of $\rm
\times 10^{37} \, erg\,s^{-1}\,Mpc^{-3}$ for the morphological classes
above.
\end{list}
\end{table*}

\begin{table*} 
\caption{Fractional contribution  of different morphological  types to
  the total AGN XLF and luminosity density}\label{tab_xlf1}
\begin{center} 
\scriptsize
\begin{tabular}{c cccc  cccc}
\hline
$\log  L_X$  &\multicolumn{4}{c}{fraction  of  $\phi_{total}(L_X)$}  &
\multicolumn{4}{c}{fraction   of   $L_X\,\phi_{total}(L_X)$}\\   ($\rm
erg\,s^{-1}$) & \multicolumn{4}{c}{}  & \multicolumn{4}{c}{} \\ & disk
& early & peculiar & point & disk & early & peculiar & point \\

41--42   &  $0.28\pm   0.12$   &  $0.30\pm0.10$   &  $0.41\pm0.22$   &
$0.001\pm0.001$ &  $0.29\pm0 0.13$  & $0.36\pm0.11$ &  $0.33\pm0.12$ &
$0.002\pm0.006$  \\   42--43  &  $0.34\pm  0.08$   &  $0.46\pm0.11$  &
$0.16\pm0.05$  & $0.02\pm0.01$  &  $0.35\pm0 0.11$  & $0.49\pm0.15$  &
$0.12\pm0.03$   &   $0.03\pm0.01$\\   43--44   &  $0.29\pm   0.10$   &
$0.47\pm0.10$  & $0.17\pm0.06$  &  $0.04\pm0.02$ &  $0.27\pm0 0.09$  &
$0.48\pm0.11$  &  $0.15\pm0.04$ &  $0.08\pm0.03$\\  44--45 &  $0.25\pm
0.11$  & $0.21\pm0.12$  &  $0.05\pm0.05$ &  $0.47\pm0.15$ &  $0.19\pm0
0.09$ & $0.15\pm0.09$ & $0.11\pm0.10$  & $0.53\pm0.20$\\ 45--46 & -- &
-- & --  & $1.00\pm1.00$ & --  & -- &  -- & $1.00\pm1.00$ \\  41--45 &
$0.30\pm0.09$   &$0.35\pm0.07$  &$0.34\pm0.15$  &   $0.010\pm0.005$  &
$0.23\pm0.05$ & $0.31\pm0.06$ & $0.13\pm0.05$ & $0.33\pm0.12$ \\

\hline
\end{tabular} 
\begin{list}{}{}
\item 
The  columns  are: (1):  X-ray  luminosity  interval; (2):  fractional
contribution  to the  XLF for  disks, early-type,  peculiar  and point
sources; (3) fractional contribution to the luminosity density for the
morphological classes above.
\end{list}
\end{center}
\end{table*}

\section{Discussion}\label{sec_discussion}

This  paper  estimates  the  space  and luminosity  density  of  X-ray
selected AGN  as a  function of their  host galaxy morphologies  in an
attempt to place constraints on the importance of major mergers, minor
interactions  and secular evolution  to the  accretion density  of the
Universe at $z\approx1$. In the standard view of major gaseous mergers
the  disks of  the participating  galaxies  are destroyed  to form  an
elliptical remnant. In this picture, disk hosts therefore indicate AGN
fueled by minor interactions or internal instabilities.

It is found  that disk galaxies represent about 25-34  per cent of the
X-ray AGN number density and  19-35 per cent of the luminosity density
of these systems (i.e.   accretion density).  These fractions are also
almost  independent of  X-ray luminosity  up to  $L_X (\rm  0.5  - 10)
\approx     10^{45}     \,      erg     \,     s^{-1}$.      Recently,
\cite{Hopkins_Hernquist2006} have developed a model for the stochastic
fueling of SMBHs via e.g.  bar instabilities or minor interactions.  In
their  model this  accretion mode  produces relatively  low luminosity
AGN, while the bulk of the  more luminous active SMBH are the result of
major  mergers.  Figure  \ref{fig_xlf} compares  our results  with the
maximal X-ray luminosity  function of the \cite{Hopkins_Hernquist2006}
stochastic accretion mode (dotted  line).  This model predicts a large
number of low luminosity AGN, which can potentially explain the entire
XLF below $\approx 10^{43}\rm  \,erg\,s^{-1}$.  Also plotted in Figure
\ref{fig_xlf}  is an  updated version  of the  XLF for  the stochastic
accretion mode  (P. F. Hopkins priv.   communication; continuous line)
in which the  AGN lightcurve decays more rapidly  than what is assumed
by \cite{Hopkins_Hernquist2006}, in  agreement with recent simulations
\citep[e.g.][]{Younger2008}  and observations  of the  Eddington ratio
distribution     of     low    redshift     AGN     in    the     SDSS
\citep[e.g.][]{Hopkins_Hernquist2008_edd}.   More  quantitatively,  in
the work  by \cite{Hopkins_Hernquist2006} the  AGN lightcurve declines
with time, $t$,  as $L \propto t^{-\beta}$ with  $\beta=0.6$, while in
the revised curve shown in  the figure $\beta=1.5$.  The updated model
reproduces the overall shape of  the observed AGN XLF of disk galaxies
but lies below the observations.   Also, both models of the stochastic
AGN  accretion mode  shown  in Figure  \ref{fig_xlf} underpredict  the
number  density  of  AGN  in  disk galaxies  at  bright  luminosities,
$>10^{44}\rm \,erg\,s^{-1}$.  This suggests that bar instabilities and
minor interactions  are more efficient in producing  powerful AGN than
the  \cite{Hopkins_Hernquist2006} model predicts.

Alternatively, one  might argue  that a fraction  of the X-ray  AGN in
disk hosts  represent major merger events.   Recent simulations indeed
show that under certain conditions disks can survive or reform after a
major merger. Some authors show that the formation of a remnant with a
large disk  requires very high cold  gas fractions ($>50$  per cent in
mass)   in   the  progenitors   \citep[e.g.][]{Springel_Hernquist2005,
  Hopkins2008},  which  may  be  typical of  high  redshift  ($z\ga2$)
objects. Others show  that cold gas flowing onto  the remnant from Mpc
scales can quickly reform a disk even if the progenitors have moderate
amounts  of cold  gas reservoirs  \citep{Governato2008}.  At  the same
time however, it  is also argued that the conditions  that lead to the
formation of a disk merger remnant are not associated with substantial
SMBH  growth  \citep[e.g.][]{Hopkins_Hernquist2008}.   The dark  matter
halo of the simulation of \cite{Governato2008} for example, has a mass
of $\rm <10^{12}\,M\odot$,  while X-ray AGN have been  shown to reside
in  halos  more  massive than  that  \citep[e.g.][]{Silverman2008_ENV,
  Gilli2008, Coil2009, Hickox2009}, where  cold gas flows are expected
to      be      suppressed      because     of      shock      heating
\citep[e.g.][]{Dekel_Birnboim2006,      Croton2006,     Cattaneo2007}.
Additionally, in the simulation of \cite{Governato2008} the black hole
mass of the merger remnant at $z=0$ is expected to be $\rm 4\times10^7
\,M\odot$, assuming  the relation between  SMBH mass and bulge  mass of
\cite{Haring_Rix2004}, while X-ray AGN  at $z\approx1$ are believed to
be   associated   with   black    holes   more   massive   than   that
\citep[e.g.]{Babic2007, Bundy2008, Coil2009, Hickox2009}.  AGN in disk
hosts therefore pose a strong  constraint on the fraction of SMBHs that
are  fueled  by mechanisms  other  than  major mergers.   Cosmological
simulations of galaxy formation and  SMBH growth will need to reproduce
the  observed luminosity distribution  of AGN  in spirals  by invoking
either  minor interactions,  bar instabilities  or the  reformation of
disks in major mergers.

If AGN in disk galaxies are systems where the accretion on the SMBH is
triggered by  secular evolution or minor tidal  disruptions then their
estimated fraction  to the  XLF represents only  a lower limit  to the
contribution of these SMBH  fueling processes to the accretion density
at $z\approx1$.   Although bulge-dominated  galaxies are likely  to be
the products of a major merger, the observed nuclear activity in these
systems may  not be associated with  this past event. The  AGN in some
early-type galaxies for  example, may be fueled by  cold gas accretion
via minor tidal  interactions. The simulations of \cite{Khalatyan2008}
demonstrate  that this  process can  activate the  SMBH  of early-type
galaxies,  thereby producing  AGN with  luminosities similar  to those
probed  here.   \cite{Ciotti_Ostriker2007}  also showed  that  secular
evolution in isolated elliptical  galaxies could drive recycled gas to
the  central regions, thereby  producing recursive  nuclear starbursts
and  accretion onto  the SMBH.  Similarly, not  all QSOs  ($L_X  > \rm
10^{44} \, erg \, s^{-1}$; point  sources in our sample) are likely to
be  the products of  major mergers.   Studies of  such sources  in the
nearby  Universe  have  revealed  that  some of  them  are  hosted  by
non-interacting   spirals  \citep[e.g.][]{Canalizo2001},   while  more
recent  observations suggest that  about 30  per cent  of QSOs  at low
redshift  are in  disk galaxies  \citep[e.g.][]{Guyon2006}. Therefore,
the fraction of AGN associated with early-type galaxies, peculiars and
point sources is  an upper limit to the  contribution of major mergers
to the accretion density.

Recent studies on star-forming galaxies also downplay the contribution
of major mergers  to the evolution of these  systems since $z\approx1$
\citep{Melbourne2005, Bell2005,Elbaz2007, Buat2008}.  It is found that
about  half  of  the   starburst  population  at  these  redshifts  is
associated   with  spirals,  underlining   the  importance   of  minor
interactions  and  bar  instabilities  in  galaxy  evolution.  If  the
build-up of  the stellar  mass of  galaxies and the  growth of  SMBH at
their centres  are related \citep[e.g.][]{Gebhardt2000, Ferrarese2000}
then  the findings  above support  the idea  that a  potentially large
fraction of  the AGN at  $z\approx1$ are triggered by  processes other
than major mergers.

\section{Acknowledgments}

The  authors  wish  to  thank  the  anonymous  referee  for  providing
constructive comments and suggestions that significantly improved this
paper and  Philip F. Hopkins for making  available unpublished results
on  the  XLF  for  the  stochastic  fuelling  of  SMBH.   The  authors
acknowledge use  of data from  the Team Keck Treasury  Redshift Survey
(TKRS;    http://www2.keck.hawaii.edu/science/tksurvey/)    and    the
ESO/GOODS project  which is based  on observations carried out  at the
Very Large Telescope at the ESO Paranal Observatory under Program IDs:
170.A-0788,  074.A-0709, 275.A-5060 and  171.A-3045. Support  for this
work  was  provided  by  NASA  through  the  Spitzer  Space  Telescope
Fellowship  Program.  The source  catalogues  used  in  the paper  are
available at www.astro.noa.gr/$\sim$age.

Based on observations obtained with MegaPrime/MegaCam, a joint project
of CFHT  and CEA/DAPNIA, at the  Canada-France-Hawaii Telescope (CFHT)
which is  operated by the  National Research Council (NRC)  of Canada,
the Institut National des Sciences de l'Univers of the Centre National
de la Recherche  Scientifique (CNRS) of France, and  the University of
Hawaii.   This work  is based  in part  on data  products  produced at
TERAPIX  and  the  Canadian  Astronomy  Data Centre  as  part  of  the
Canada-France-Hawaii Telescope Legacy  Survey, a collaborative project
of NRC and CNRS.

\bibliography{mybib}{}
\bibliographystyle{mn2e}

\end{document}